\documentclass{article}
\newcommand{\ignore}[1]{} 
\usepackage{spconf,amsmath,graphicx,multirow, hyperref}
\title{Improved Multipath Time delay estimation using cepstrum subtraction}
\twoauthors
  {\begin{tabular}{c}Eric L. Ferguson\sthanks{The author gratefully acknowledges the contributions provided by the IEEE Oceanic Society Scholarships, the University of Sydney Electrical and Information Engineering PhD Scholarship, and the Defence Science and Technology Group Postgraduate Research Scholarship. }, Stefan B. Williams\end{tabular}}
	{Australian Centre for Field Robotics\\
		 The University of Sydney, Australia}
  {Craig T. Jin}
	{Computing and Audio Research Laboratory\\
		 The University of Sydney, Australia}

\begin{document}
\ninept
\maketitle
\begin{abstract}
When a motor-powered vessel travels past a fixed hydrophone in a multipath environment, a Lloyd's mirror constructive/destructive interference pattern is observed in the output spectrogram.
The power cepstrum detects the periodic structure of the Lloyd's mirror pattern by generating a sequence of pulses (rahmonics) located at the fundamental quefrency (periodic time) and its multiples.
This sequence is referred to here as the `rahmonic component' of the power cepstrum.
The fundamental quefrency, which is the reciprocal of the frequency difference between adjacent  interference fringes, equates to the multipath time delay. 
The other component of the power cepstrum is the non-rahmonic (extraneous) component, which combines with the rahmonic component to form the (total) power cepstrum.
A data processing technique, termed `cepstrum subtraction', is described.
This technique suppresses the extraneous component of the power cepstrum, leaving the rahmonic component that contains the desired multipath time delay information.
This technique is applied to real acoustic recordings of motor-vessel transits in a shallow water environment, where the broadband noise radiated by the vessel arrives at the hydrophone via a direct ray path and a time-delayed multipath.
The results show that cepstrum subtraction improves multipath time delay estimation by a factor of two for the at-sea experiment.
\ignore{
In a shallow water environment, when a broadband source of radiated noise transits past a fixed hydrophone, a Lloyd's mirror constructive/destructive interference pattern can be observed in the output spectrogram. 
The fundamental quefrency, which is the reciprocal of the frequency difference between adjacent  interference fringes, equates to the multipath delay time. 
By taking the spectrum of a (log) spectrum, the power cepstrum detects the periodic structure of the Lloyd's mirror fringe pattern by generating a sequence of pulses located at the fundamental quefrency and its multiples.
A novel data processing method is introduced that supresses cepstrum features extraneous to rhamonics (time delays) in order to improve their detection, and is termed `cepstrum subtraction'.
An experiment is conducted where a motor-powered boat transits past a hydrophone located $1$~m above the seafloor in shallow water ($20$~m deep), and the time delay between with the direct propagation ray path and seafloor reflected multipath is estimated.
Time delay estimation performance is compared with other single-sensor time delay estimation methods.
Results show improved time delay estimation performance in the shallow water experiment, and improved robustness to poor signal-to-noise levels.}
\end{abstract}
\section{Introduction}\label{sec:intro}
In underwater acoustics, the Lloyd's mirror effect~\cite{urick1975principles, carey2009lloyd} is observed as an interference pattern in the spectrogram (variation with time of the spectrum) of a receiving hydrophone.
It occurs when the sinusoidal signal components that compose a source of broadband radiated noise arrive at the receiver via the direct propagation path and combine with amplitude-scaled time-delayed replicas arriving via indirect propagation paths (multipaths). 
If the two signal arrivals are \textit{in phase}, they reinforce each other through coherent addition resulting in \textit{constructive} interference; when they are in \textit{antiphase}, they cancel and \textit{destructive} interference occurs.
As a result, the received intensities of a source's spectral lines are observed to undergo a periodic, or cyclic, variation with range~\cite{kapolka2008equivalence}.\ignore{ 
For a broadband source composed of a series of sinusoidal signals with different frequencies, the variation with range of the received spectrum is referred to in the literature as ``range-frequency striation", which can be understood by examining the nature of phase coherent addition of signals arriving via propagation paths of different lengths~\cite{jones2012application}.
Similarly, when a broadband source transits past a fixed sensor, the destructive interference fringes appear as a set of hyperbolic loci in the received spectrogram, with the vertices uniformly spaced in frequency along a vertical line corresponding to the time at which the source is at the closest point of approach to the sensor~\cite{kapolka2008equivalence}.

The interference pattern is attributed to the signal arriving at the receiver via the \textit{direct} sound propagation ray path combining with the signal arriving via an \textit{indirect} propagation ray path that involved a reflection from the seafloor boundary.}
The resultant intensity is given by
\begin{align}
A_r^2 = A_d^2 + A_i^2 + 2 A_d A_i \cos(\kappa \delta l) \text{ , } \label{eq:resultant_intensity}
\end{align}
where $A_d$ and $A_i$ are the amplitudes of the respective direct path and indirect path arrivals, $\kappa = 2 \pi / \lambda$ ($\lambda$ denotes wavelength) is the spatial frequency (or wavenumber) and $\delta l$ is the path length difference between the two paths; the last term in \eqref{eq:resultant_intensity} is oscillatory due to interference between the direct path and multipath arrivals, which gives rise to a Lloyd's mirror interference pattern~\cite{jones2012application}.
Destructive interference fringes occur when $\kappa \delta l = (2 k - 1)\pi \text{ , } k=1,2,3,...,$ where $k$ is the fringe order (number).
For adjacent destructive interference fringes
\begin{align}
(\kappa_{k+1} - \kappa_k)\delta l = 2\pi \text{ . } \label{eq:freq_range_between_two_striations}
\end{align}
Hence from \eqref{eq:freq_range_between_two_striations}, at any given instant of time, the destructive interference fringes have a uniform spacing in the direction of the frequency axis which is given by
\begin{align}
(f_{k+1} - f_k ) = \frac{c}{\delta l} = \frac{1}{\tau_{i,d}}\text{ , } 
\end{align}
where $c$ is the isospeed of sound travel in the underwater medium and $\tau_{i,d}$ is the time difference in the arrival times of the signal propagating via the respective indirect and direct paths.
Note that the temporal frequency difference $f_{k+1} - f_k $ equates to the reciprocal of the multipath time delay.

\ignore{Additionally, the multipath delay (which depends on the range of the source) is given by~\cite{ferguson2016deep}
\begin{align}
\tau_{i,d} = \frac{\sqrt{D^2 + a^2}-{\sqrt{D^2 + b^2}}}{c} \text{ , } \label{eq:passive_ground_range}
\end{align}
where $D$ is the ground range (the horizontal distance between the source and the sensor), $a$ and $b$ are constants, i.e. $a=h_s+h_r = 21$~m, $b=h_s-h_r = 19$~m, given that the altitude of the source $h_s=20$~m and the altitude of the hydrophone $h_r=1$~m.
Hence, a single-sensor spectrogram displaying a Lloyd's mirror interference pattern has periodic structure which contains source localization (range) information.
If the water depth, sensor altitude, and isospeed of sound travel in the underwater medium are known, then knowledge of the  multipath time delay enables localization of a source on the sea surface by estimating its range from a single omnidirectional sensor~\cite{ferguson2016deep}.}

Fundamentally, the \textit{cepstrum} is the spectrum of a logarithmic spectrum and it is used in practice to detect \textit{periodicity in a frequency spectrum}. The \textit{quefrency} is a measure of the periodicity of the spectral ripple and it has the units of (periodic) time.
In speech analysis, periodicity in the voice spectrum is observed as a regular spacing in frequency of the fundamental voice frequency (pitch) and its harmonics~\cite{noll1964short, noll1967cepstrum, oppenheim2004frequency}.
If the spectrum of the logarithm of the speech power spectrum is computed, a peak will appear corresponding to the pitch period of the voiced-speech segment being analysed~\cite{noll1967cepstrum}.

For the present case of a transit of a broadband source past an acoustic sensor in an underwater acoustic multipath propagation environment, periodic structure is observed in the output spectrogram as an ordered arrangement of Lloyd's mirror interference fringes which form a regular periodic pattern. 
The \textit{quefrency} can be considered as the periodic time associated with a series of interference fringes that are uniformly spaced in frequency.
The power cepstrum is generated by taking the inverse Fourier transform of the logarithm of the power spectrum~\cite{randall2017history}.
The first rahmonic is the fundamental period, which is equal to the reciprocal of the frequency interval between any two adjacent destructive interference fringes (or, equally, between any two adjacent constructive interference fringes).\ignore{
Periodic time (or quefrency) depends on the distance of the source from the sensor.
The quefrency increases as the source closes in range and reaches a maximum when the source is at the closest point of approach to the sensor, after which the quefrency decreases as the source opens in range.}
Hence, cepstrum analysis is suited to detecting and quantifying periodicity in a spectrogram. It is also suited to multipath time delay estimation~\cite{ferguson2017convolutional, ferguson2018sound}.

The original contribution of this research is a novel data processing technique that suppresses cepstrum components extraneous to the rahmonics, thereby improving the estimation of the multipath time delay for broadband signals arriving at a receiving hydrophone.
This improvement in time delay estimation is demonstrated using real data collected during an at-sea experiment. \ignore{ 
It should be noted that by estimating the time delay between a direct propagation path and a seafloor reflected multipath allows the source on the sea-surface to be passively localized in range using a single sensor~\cite{ferguson2016deep}. This directly improves passive range estimation of transiting a motor-powered boats in a very shallow water environment.}

\section{Underwater Acoustic Sensor Output Model}\label{sec:2}

\subsection{Direct path signal combined with indirect path signal}
For the present case, the signal of interest is continuous broadband acoustic noise radiated by a small motor-powered boat underway.
The output of the hydrophone, which is located above the seafloor, is modelled as a combination of: (a) an underwater acoustic signal propagating directly from the source to the sensor $s(t)$, where $t$ denotes time, and (b) an amplitude-scaled time-delayed replica propagating via an indirect path that includes a seafloor boundary reflection $\alpha s(t-\tau_\beta) $, where $\alpha$ is the attenuation (or boundary reflection) coefficient ($0<\alpha\leq1$), and $\tau_{\beta} $ is the time delay.
The combination of the signal with a single echo can be represented as~\cite{oppenheim2004frequency}
\begin{align}
x(t) = s(t) + \alpha s(t-\tau_\beta) . \label{eq:simple_signal_equation}
\end{align}

The time delay between the indirect propagation path and the direct propagation path results in a phase difference between the direct path and multipath signals equivalent to $2 \pi f \tau_\beta$, where $f$ is the signal frequency.
Assuming that the transmission loss is the same for each propagation path, the instantaneous power of the resultant signal is given by~\cite{oppenheim2004frequency}
\begin{align}
\big|X(f)\big|^2 = \big|S(f)\big|^2  \big( 1 + \alpha^2 +2\alpha \cos(2 \pi f \tau_\beta) \big) \text{ ,}\label{eq:instantaneous_power_direct_and_indirect_b}
\end{align}
where the power spectrum of the direct path signal  $\big|S(f)^2\big| $ is modulated by a periodic function of the frequency.
Taking the logarithm of the output power spectrum converts the \textit{product} in \eqref{eq:instantaneous_power_direct_and_indirect_b} to a \textit{sum} of two components, i.e.
\begin{footnotesize}
\begin{equation}
\begin{aligned}
\log\big|X(f)\big|^2 =& \log\big|S(f)\big|^2 + \log  \big( 1 + \alpha^2 +2\alpha \cos(2 \pi f \tau_\beta) \big) \text{,} \label{eq:instantaneous_log_power_direct_and_indirect}
\end{aligned}
\end{equation}
\end{footnotesize}where the additive periodic component has a period determined by the multipath delay $\tau_{\beta}$.

\subsection{Power cepstrum}
For the periodic structure of a Lloyd's mirror interference pattern, the quefrency (or periodic time) of the first harmonic is the reciprocal of the uniform frequency difference between adjacent destructive interference fringe minima\ignore{ (or constructive interference fringe maxima)}. 
For the present case of Lloyd's mirror interference pattern formation resulting from the combination of the direct and indirect propagation path signals at the sensor, the quefrency of the first rahmonic is the multipath time delay.
In other words, cepstrum analysis of the hydrophone's noisy signal output provides an estimate of the multipath time delay $\tau_\beta$, which is the difference in the arrival times of the direct path signal and the indirect path signal  at the hydrophone.
The definition of the \textit{power cepstrum} $C_{XX}(\tau)$ can be expressed mathematically as~\cite{randall2017history}
\begin{equation}
\begin{aligned}
C_{XX}(\tau) \ignore{&= \mathcal{F}^{-1} \big\{ \log S_{XX}(f) \big\}} = \mathcal{F}^{-1} \big\{ \log \big|X(f)\big|^2 \big\}\text{ ,} 
\end{aligned}
\end{equation}
where $\tau$ is the quefrency (or delay time), $\mathcal{F}^{-1}$ denotes the inverse Fourier transform and $\big|X(f)\big|^2$ is equivalent the two-sided power spectrum of a received signal $x(t)$.

In the present case, the power cepstrum is given by the inverse Fourier transform $\mathcal{F}^{-1}$ of \eqref{eq:instantaneous_log_power_direct_and_indirect}, which can be written as
\begin{align}
C(\tau) = \mathcal{F}^{-1} \big(\log\big|S(f)\big|^2\big) + \sum^{\infty}_{n=1} a_n \delta(\tau - n \tau_{\beta}) \text{ .}  \label{eq:simple_power_cepstrum}
\end{align}
where $a_n$ is the strength of the $n$\textsuperscript{th} rahmonic and it is given by 
\begin{align}
a_n = \frac{2}{n}(-1)^{n+1} \alpha ^n \text{ for } n = 1,2,3, ... \label{eq:Fourier_coefficients}
\end{align} 
The rahmonic strengths are observed to alternate in sign and decrease as the reciprocal of the rahmonic number $n$ in the same way as the Mercator series.
Hence, the power cepstrum of the received signal consists of the power cepstrum of the direct path signal $S(f)$ and a sequence of impulse functions (or rahmonics) of strength $a_n$ located at quefrency $\tau_{\beta}$ and its multiples.
\hyperref[fig:cepst_components]{Figure 1(a)} shows the computed power cepstrum $C(\tau)$ when a  broadband direct-path signal is combined with a time-delayed replica ($\alpha=1$, $\tau_{\beta}=224$~$\mu$s).
\hyperref[fig:cepst_components]{Figure 1(b)}  and \hyperref[fig:cepst_components]{Figure 1(c)} show the variation with quefrency of the respective power cepstrum components; the non-rahmonic component which is given by $C_1(\tau) = \mathcal{F}^{-1}\big( \log \big |S(f)\big|^2\big)$ and the rahmonic impulse component which is given by
$C_2(\tau) = \sum^{\infty}_{n=1} a_n \delta(\tau-n\tau_{\beta})$.
\hyperref[fig:cepst_components]{Figure 1(d)} shows the autocorrelation as a function of positive lag with the maximum value of the echo's sinc function corresponding to a lag (i.e. the echo time delay) of $224$~$\mu$s.
In practice, additive noise biases the autocorrelation function and may obscure the peaks corresponding to propagation time delay difference~\cite{bendat1980engineering}.
Also, when there is more that one indirect propagation path, the identification of individual propagation paths using autocorrelation analysis becomes more difficult, and so cepstrum analysis is preferred because each propagation path has a series of more readily identifiable cepstrum peaks located at the rahmonic quefrencies that characterize a specific propagation path~\cite{bendat1980engineering}.

\subsection{Cepstrum subtraction}\label{sec:cepst_subtract}
\begin{figure}[t]
  \centering
\includegraphics[width=1.0\linewidth]{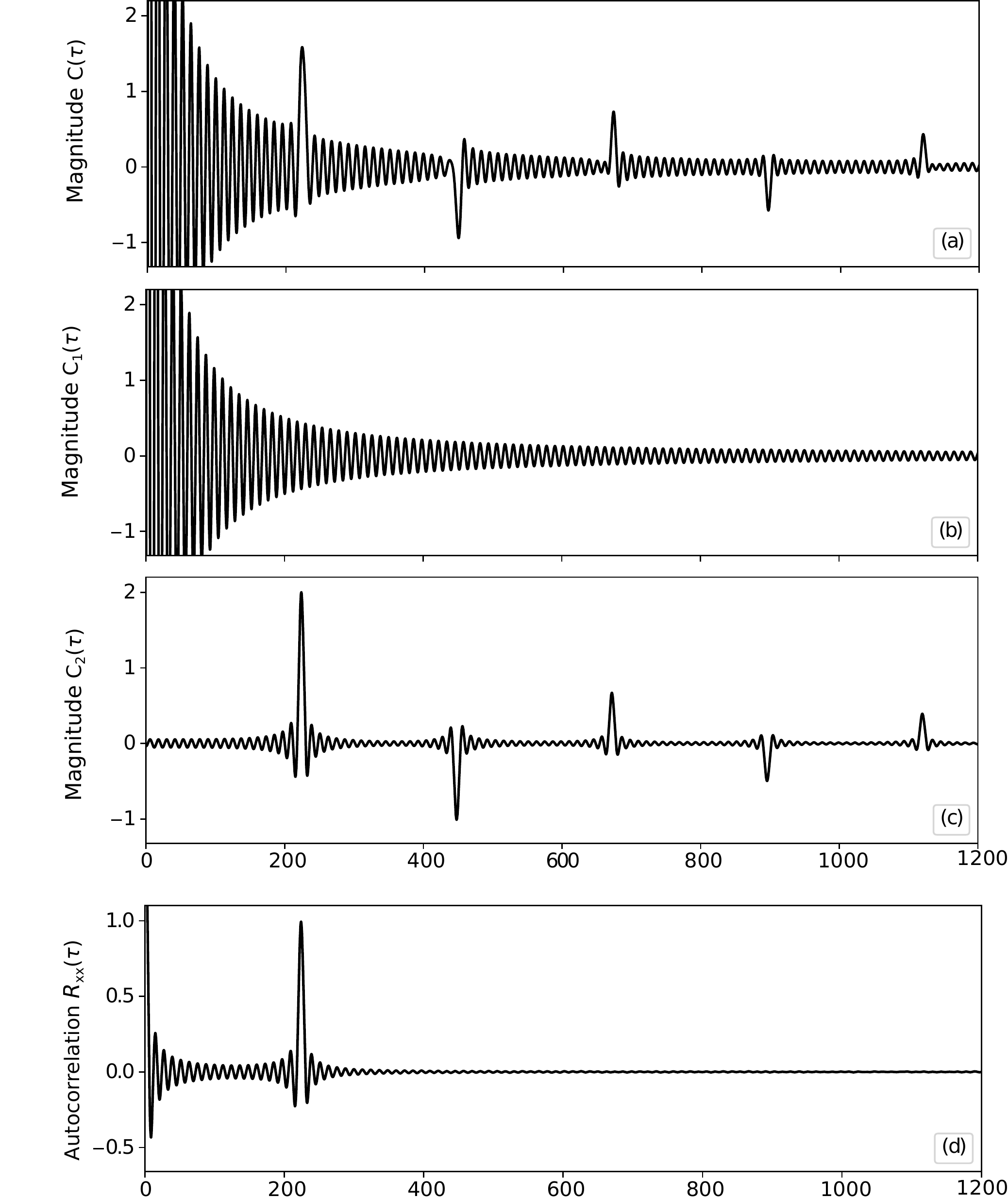}
 \caption{(a) Variation with quefrency of the power cepstrum $C(\tau)$ of a direct-path signal combined with an indirect-path echo $\alpha=1$, $\tau_{\beta}=224$~$\mu$s. (b) Variation with quefrency of the power cepstrum component $C_1(\tau)$. (c) Similar to (b), but for the rahmonic impulse component $C_2(\tau)$. (d) Variation with lag of the autocorrelation function for the direct path signal combined with the multipath echo.}
\label{fig:cepst_components}
\end{figure}

A novel data processing technique is presented that suppresses cepstrum components extraneous to the rahmonics.
From \eqref{eq:simple_power_cepstrum}, the power cepstrum of the received signal $C(\tau)$  consists of the non-rahmonic components $C_1(\tau)$ derived from stationary process $S(f)$ (which remains stationary after inverse Fourier transformation~\cite{bendat1971random_pg56to98}), and a sequence of rahmonic impulse functions $C_2(\tau)$ for another (independent) stationary process~\cite{bendat1971random_pg1to36}. Note that $C_1(\tau)$ is independent of the time delay variable  $\tau_{\beta}$, whereas $C_2(\tau)$  is dependent with impulses located at quefrency $\tau_{\beta}$ and its multiples. 

The ensemble average of the power cepstrum over $M$ realizations may be given by
\begin{align}
\frac{1}{M} \sum^M_{m=1} C_{m}(k)  = \frac{1}{M} \sum^M_{m=1} C_{1,m}(k) + \frac{1}{M} \sum^M_{m=1} C_{2,m}(k)   \text{ , } \label{eq:mean_cepstra_a} 
\end{align}
where $C_{1,m}(k)$ and $C_{2,m}(k)$ are the $m$\textsuperscript{th} realizations at discrete quefrency $k$ for the non-rahmonic and rahmonic components, respectively.
For $\tau_{\beta} \neq 0$ and assuming $\tau_{\beta}$ is independent of realization $m$, then the expected value of $C_{2}(k)$ at any quefrency\ignore{ is given by the mean of the Dirac (impulse) function which} is zero, i.e.
\begin{align}
 \lim_{M \to \infty} \frac{1}{M} \sum^M_{m=1} C_{2,m}(k) = 0  \text{ .} \label{eq:limit_dirac} 
\end{align}
Therefore, the non-rahmonic component $C_{1}(k)$ can be estimated by taking the average of the received power cepstrum $\bar{C}_{1}(k) $ over a sufficiently long time (i.e. for a sufficiently large $M$),  where
\begin{align}
\bar{C}_{1}(k) = \frac{1}{M} \sum^M_{l=1} C_{m}(k)  \text{ . } \label{eq:mean_cepstra} 
\end{align}
\ignore{From \eqref{eq:limit_dirac}, }Note that the long time-averaged rahmonic contribution at quefrency $k$ is negligible for a sufficiently large $M$.
This approach is consistent with other ensemble averaging methods~\cite{bendat1971random}.

The rahmonic component for realization $m$ can be estimated by subtracting the estimate of the non-rahmonic component from the received power cepstrum.
The cepstrum subtraction process is described mathematically by,
\begin{align}
\hat{C}_{2,m}(k) = C_{m}(k) - a(k)\bar{C}_{1}(k) \text{ ,} \label{eq:cepstrum_subtraction}
\end{align}
where $ \hat{C}_{2,m}(k)$ is an estimate of the rahmonic component for the $m$\textsuperscript{th} frame at discrete quefrency $k$, $C_{m}(k)$ is the received power cepstrum for the $m$\textsuperscript{th} frame at quefrency $k$, and  $\bar{C}_{1}(k)$ is the long time-averaged received cepstrum given by \eqref{eq:mean_cepstra}.\ignore{It is assumed that $\bar{C}_{1}(k)$ is derived from a stationary random process $S(f)$.}
The parameter $a(k)$ controls the amount of the non-rahmonic component that is subtracted from the power cepstrum. 
For full non-rahmonic component subtraction $a(k) = 1$ and for over-subtraction $a(k) > 1$, analogous to spectral subtraction~\cite{vaseghi2008advanced}.
The cepstrum subtraction process is visualized in \autoref{fig:cepst_components}, where the non-rahmonic contribution in (b) is subtracted from the received power cepstrum (a), leaving only the rahmonic component in (c).

Note that the cepstrum subtraction method has the same formulation as cepstrum mean normalization, which is used in speech processing, but with the addition of the factor $a(k)$~\cite{noll1964short, garner2011cepstral, furui1981cepstral}.
Rather than removing channel-effects (distortions) for Mel-frequency cepstrum coefficients, instead cepstrum components extraneous to the rahmonics are suppressed. 
\ignore{The methods are applied to different signal models.}Also the two methods are applied to the outputs of different signal models.

\begin{figure}[t]
  \centering
 \includegraphics[width=1.0\linewidth]{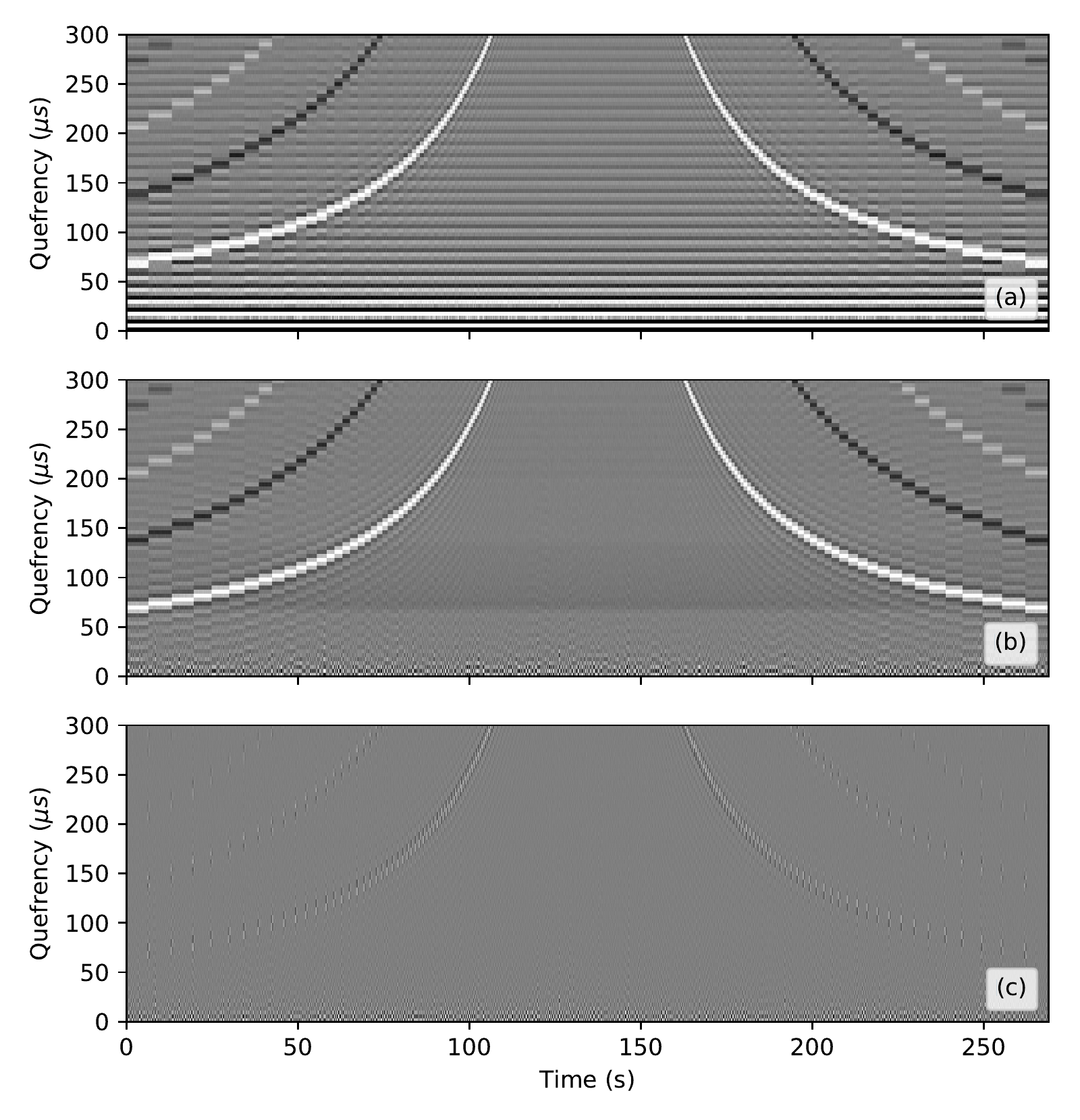}
 \caption{Various power cepstrograms for simulated acoustic data, with quefrencies up to $300$~$\mu$s. (a) Variation with time of the power cepstrum, (b) Similar to (a), but  with cepstrum subtraction where the non-rahmonic component is estimated from the entire transit ($M=2700$), and (c) Similar to (b) but the non-rahmonic component is estimated from the last $2$~seconds ($M=20$) of received cepstra,  which have significant rahmonic contributions that bias the estimate. \label{fig:CMN_cepstrogram_comparison}}
\end{figure}

\autoref{fig:CMN_cepstrogram_comparison} shows various power cepstrograms for a simulated  motor-boat transit using synthetic broadband radiated noise.
The source/sensor geometry is the same for the at-sea experiment discussed below in \autoref{sec:data}.
\autoref{fig:CMN_cepstrogram_comparison}(a) shows the power cepstrum without cepstrum subtraction.
The horizontal banding associated with the non-rahmonic component is prominent, which can be seen to interfere with the rahmonics.
\autoref{fig:CMN_cepstrogram_comparison}(b) shows that the effect of cepstrum subtraction is to suppress the non-rahmonic component, thus restoring the fidelity of the rahmonic component.
Here, $M=2700$ which is sufficiently large so that the contribution of the rahmonic component to the time averaged-power cepstrum is negligible. 
The strengths of the rahmonics relative to the non-rahmonics have increased.
In contrast, \autoref{fig:CMN_cepstrogram_comparison}(c) shows that the cepstrum subtraction process has suppressed both the rahmonic and non-rahmonic components because $M=20$ is small, i.e. the contribution of the rahmonic component to the time-average power cepstrum is significiant.
Cepstrum subtraction requires then multipath time delay $\tau_{\beta}$ to be a rapidly changing function of $m$, otherwise the rhamonic strengths bias the estimate of the non-rahmonic component of the power cepstrum, violating  \eqref{eq:limit_dirac}.
\ignore{
In \autoref{sec:cepst_subtract}, it is assumed that the time delay $\tau_{\beta}$ is independent of the received cepstrum realizations. 
When this assumption is violated, the estimate of the non-rahmonic components becomes biased since the rahmonic component in \eqref{eq:limit_dirac} no longer is zero.
The non-rahmonic estimate in (c) is biased due to the similar time-delay values over the cepstrum realizations, resulting in the suppression of non-rahmonic and rahmonic components alike.
The non-rahmonic estimate in (b) is less biased due to a wider range of time-delay values over the cepstrum realizations, and rahmonics are preserved.}
  
\section{Experimental Results}\label{sec:experiment}

\begin{figure}[t]
  \centering  
\includegraphics[width=1.0\linewidth]{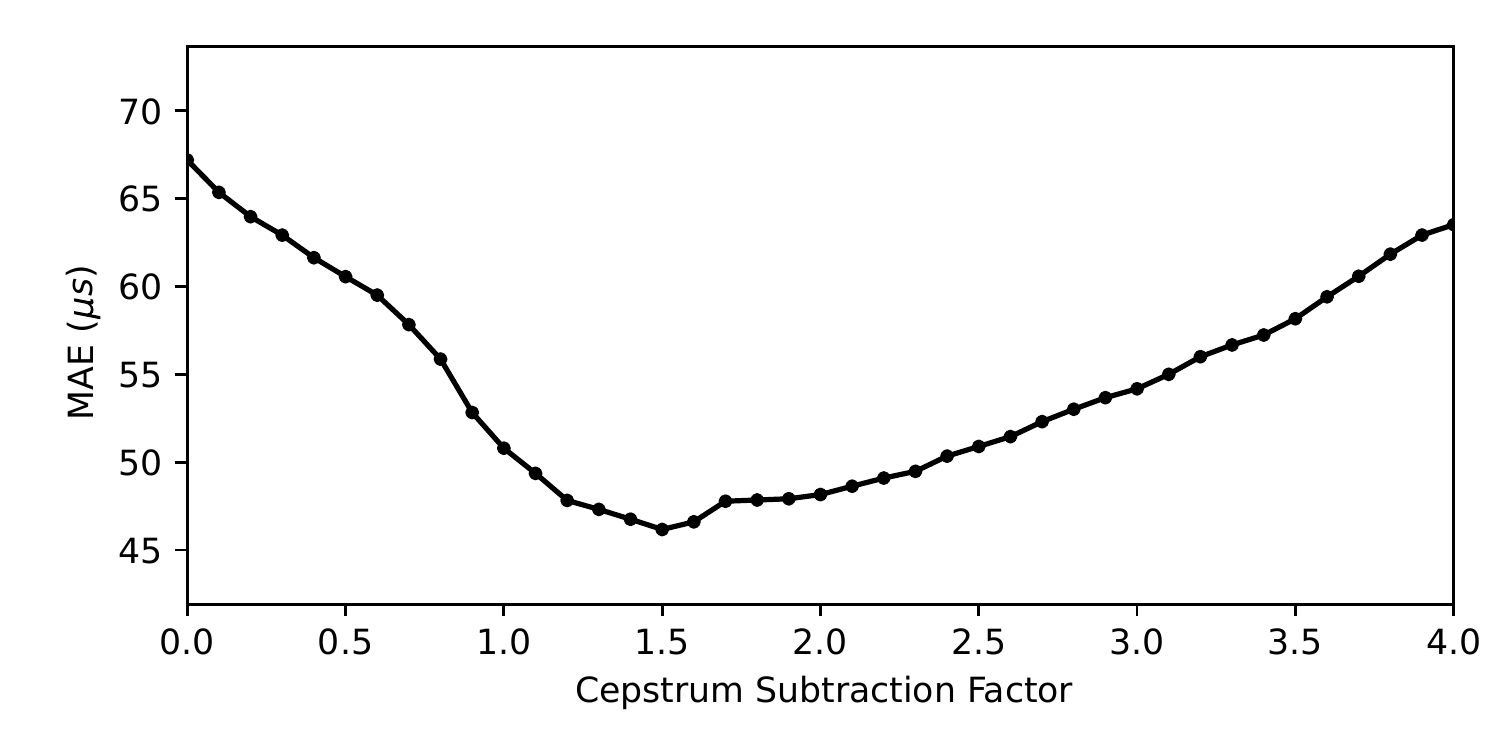}
 \caption{Variation with the cepstrum subtraction factor $a(k)$ of the mean absolute error for multipath time delay estimation. \label{fig:cepst_oversubtraction_results}} 
\includegraphics[width=1.0\linewidth]{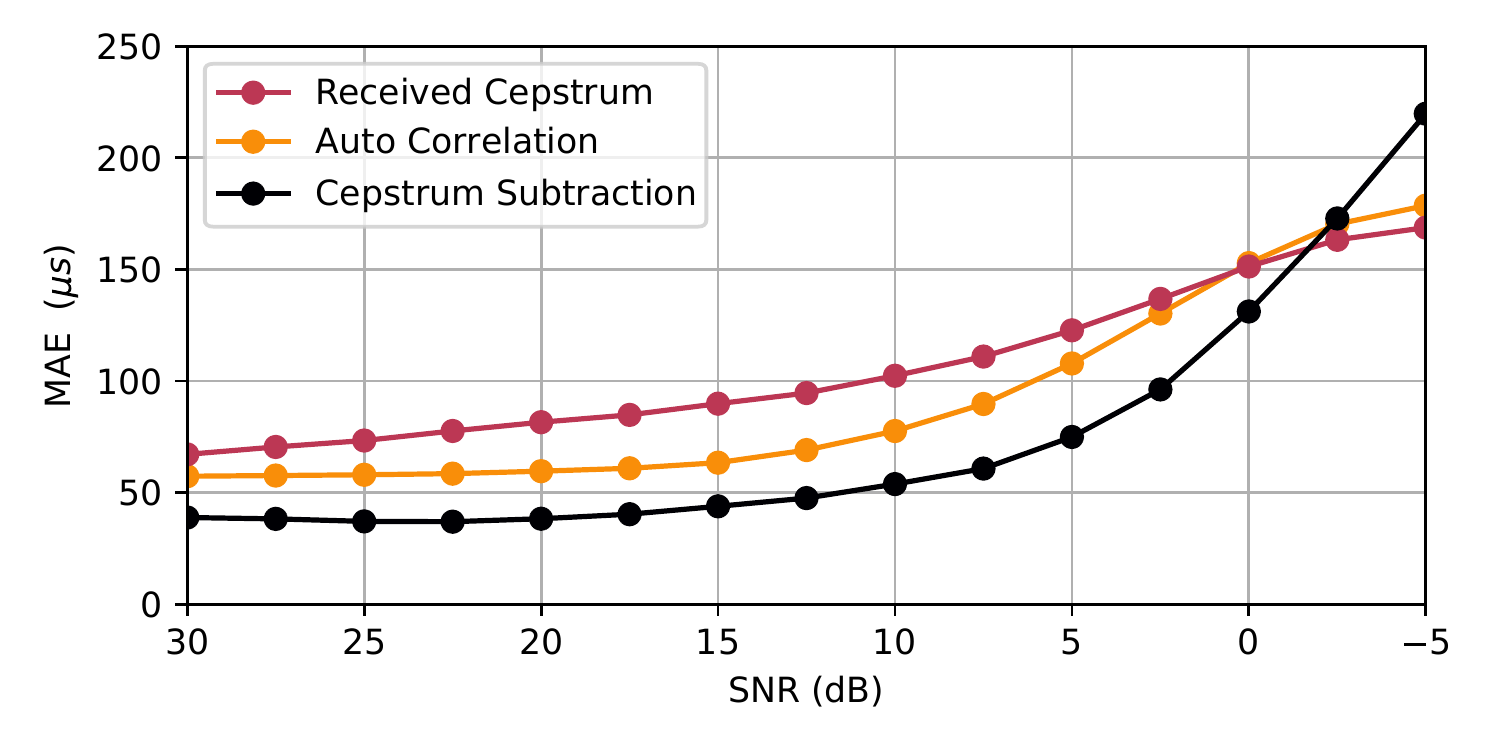}
 \caption{Variation with signal-to-noise ratio of the mean absolute error for mulitpath time delay estimation. \label{fig:TDE_vs_SNR}}
\end{figure}

The multipath time delay, or the time difference in the arrivals of the signal via the direct propagation path and the signal via the indirect seafloor-reflected multipath, is estimated using various cepstrum methods and the autocorrelation function. 
The performance  of the time delay estimation methods are compared using acoustic data collected during an at-sea experiment.

\subsection{Dataset}\label{sec:data}
An experiment was conducted where a propeller-driven boat, which radiated broadband underwater acoustic noise, transited past a hydrophone located  $1$~m above the seafloor in very shallow water ($20$~m deep). 
Fourteen transits were recorded over a two day period, where the vessel approached the sensor from different directions.
A spectrogram of the hydrophone's output was observed to display a Lloyd's mirror interference pattern over the entire frequency band ($\leq$$90$~kHz) of the receiver, which was sampled at $250$~kHz.
Recording commenced when the vessel was inbound at a ground range of $200$~m.
The vessel transited past the sensor, and recording was terminated when the vessel was $200$~m outbound.
The vessel's position was logged\ignore{ relative to the receiving hydrophone} at $0.1$~s intervals using a RF tracking device. 

The multipath time delay is predicted as a function of the vessel's ground range, assuming an isovelocity ($1520$~ms$^{-1}$) sound propagation medium and specular reflection from a flat seafloor~\cite{ferguson2016deep}. 
Acoustic recordings are divided into $0.1$~s duration sound clips. 
The power cepstrum, power cepstrum with cepstrum subtraction, and autocorrelation function are computed for each sound clip.
The estimate of the multipath time delay is the quefrency (or lag) corresponding to the  peak value in each of these functions.
The estimated multipath time delay   is compared with the predicted time delay.
Six minutes of background recordings (with no source present) was also collected in order to measure the ambient noise spectrum, which was used to produce various levels of additive noise for signal-to-noise ratio (SNR) experiments.

\subsection{Comparison of time delay estimation methods}

\autoref{fig:cepst_oversubtraction_results} shows the mean absolute error (MAE) for multipath time delay estimation with cepstrum subtraction, as a function of the subtraction factor $a(k)$ in \eqref{eq:cepstrum_subtraction}. When $a(k) = 0$, there is no cepstrum subtraction and the time delay is estimated from the received power cepstrum. When $a(k) = 1$, full cepstrum subtraction occurs and  when $a(k) > 1$, there is over-subtraction.
Non-rahmonic components are estimated by taking the mean of the received power cepstra for the entire acoustic dataset.
The MAE decreases with cepstrum subtraction, which demonstrates the  benefit of non-rahmonic component suppression in the received power cepstrum.
The minimum MAE occurs when $a(k) = 1.5$.
This level of over-subtraction reduces the MAE by $20$~$\mu$s when compared with nil subtraction i.e. $a(k)  = 0$.

The multipath time delay is estimated from (a) the power cepstrum, (b) the power cepstrum with cepstrum (over) subtraction, and (c) the autocorrelation function. 
The estimate of the multipath time delay is the quefrency (or lag) corresponding to the peak value in each of these functions.
Colored noise with the same power spectral density as background noise recordings (without the source) is added to recordings in varying amounts in order to change the SNR. 
This enables time delay estimation performance to be tested as a function of SNR.\ignore{
The cepstrum-based time delay estimation methods are shown to be robust to changes in the signal-to-noise levels of recordings, when compared with the results from autocorrelation.
The power cepstrum with cepstrum subtraction provides the best overall performance, showing that suppression of non-rahmonic components can lead to improved time delay estimation performance. }
\autoref{fig:TDE_vs_SNR} shows that cepstrum subtraction provides the overall performance, confirming that suppression of non-rahmonic components improves time delay estimation performance.

\section{Conclusion}\label{sec:concl}
Cepstrum analysis enables estimation of the multipath time delay for broadband signals arriving via direct and indirect signal propagation paths at a receiving hydrophone.
A cepstrum subtraction method suppresses the cepstrum components extraneous to the rhamonics.
As a result, the magnitudes of the rahmonics relative to the non-rahmonics are increased.
Also, there is an improvement in multipath time delay estimation, which is demonstrated using real data collected during an at-sea experiment. 
The experimental results showed improved multipath time delay estimation performance in a very shallow water environment when compared with other time delay estimation methods, and robustness to decreasing signal-to-noise ratios.
Multipath time delay estimation using cepstrum subtraction is superior to using the autocorrelation function when the SNR $\geq 0$.
However, without cepstrum subtraction, the autocorrelation function is better than the power cepstrum for multipath time delay estimation in the present application.
Cepstrum over-subtraction improves multipath time delay estimation.

\bibliographystyle{IEEE}
\bibliography{references.bib}

\end{document}